\documentclass[conference]{IEEEtran}
\usepackage{graphicx}
\usepackage{epstopdf} 
\usepackage{verbatim} 
\usepackage{multirow} 
\usepackage{multicol}
\usepackage{amsfonts}
\usepackage{bbm}
\usepackage{subfigure}
\usepackage{colortbl} 
\usepackage{indentfirst}
\usepackage{dcolumn,booktabs}
\usepackage{algorithm}
\usepackage[noend]{algpseudocode}
\usepackage{algorithmicx,algorithm}
\usepackage{setspace}
\usepackage{algpseudocode}
\usepackage{graphics}
\usepackage{epsfig,color}
\usepackage{amsmath}
\usepackage{amssymb}
\usepackage[english]{babel}
\ifCLASSINFOpdf

\else
\fi
\newtheorem{Theorem}{Theorem}

\newtheorem{Lemma}{Lemma}

\newtheorem{Remark}{Remark}

\begin{document}
\title{Spatio-Temporal Analysis of SINR Meta Distribution for mmWave Heterogeneous Networks Under Geo/G/1 Queues}

\author{
\IEEEauthorblockN{Le Yang, Fu-Chun Zheng and Shi Jin}
}
\maketitle

\begin{abstract}
A fine-grained analysis of network performance is crucial for system design. In this paper, we focus on the meta distribution of the signal-to-interference-plus-noise-ratio (SINR) in the mmWave heterogeneous networks where the base stations (BS) in each tier are modeled as a Poisson point process (PPP). By utilizing stochastic geometry and queueing theory, we characterize the spatial and temporal randomness while the special characteristics of mmWave communications, including different path loss laws for line-of-sight and non-line-of-sight links and directional beamforming, are incorporated into the analysis. We derive the moments of the conditional successful transmission probability (STP). By taking the temporal random arrival of traffic into consideration, an equation is formulated to derive the meta distribution and the meta distribution can be obtained in a recursive manner. The numerical results reveal the impact of the key network parameters, such as the SINR threshold and the blockage parameter, on the network performance.
\end{abstract}


%
\IEEEpeerreviewmaketitle

\section{Introduction}
\subsection{Motivation}
With the rapid proliferation of the communications and electronic technologies, various new applications have emerged, such as autonomous vehicles, virtual reality and augmented reality. As a result, the fifth generation (5G) and beyond 5G networks are anticipated to provide massive connectivity for massive amount of services \cite{massive-connectivity}. Specifically, the data rate demand of virtual and augmented reality can reach the order of several gigabits-per-second or higher \cite{AR-VR}. Under this circumstance, millimeter wave (mmWave) communications stands out as a promising approach to meet the increasing data rate demand \cite{mmwave-technology}.

The performance of the mmWave networks has been studied by a variety of research works. In \cite{K-tier-Renzo}, the authors proposed a generalized mathematical framework for the analysis of the mmWave heterogeneous networks and the coverage probability and the average rate were obtained. In \cite{Bai}, a distance-dependent line-of-sight (LOS) probability function was utilized in cellular networks where the LOS and non-line-of-sight (NLOS) Base Stations (BSs) were distributed as two independent non-homogeneous Poisson point processes (PPP). In \cite{K-tier}, the authors obtained the expressions of the coverage probability and energy efficiency in the mmWave heterogeneous networks by utilizing the multi-ball approximation for the blockage model.

However, each link is assumed to have a full buffer in the above works \cite{JSAC}, i.e., there is always a packet to be transmitted. This assumption restricts the application of these works under various traffic conditions. To overcome this drawback, an additional dimension of randomness is introduced by considering the temporal domain, i.e., the arrival and service of the packets at a user is considered as a queueing process \cite{1}-\cite{3}. However, the effect of queue status cannot be characterized explicitly in these works. Fortunately, the notion of meta distribution is introduced in \cite{md-maenggi} and can serve as a powerful tool to characterize the service intensity for each individual queue. The conditional STP is defined as the probability of the signal-to-interference-plus-noise-ratio (SINR) exceeding a pre-defined threshold $\theta$, i.e., $\mathbbm{P}(\text{SINR}>\theta|\Phi)$, given a spatial realization of BSs and users. The STP is obtained by averaging the conditional STP (a random variable) over the BS distributions and channel fading, and therefore cannot reflect the SINR variation among the individual links. In other words, the STP only answers the question of ``On average what fraction of users experiences successful transmission (i.e., $\text{SINR}>\theta$)?''. To overcome this drawback and obtain a fine-grained analysis on the SINR distribution of the mmWave heterogeneous networks and inspired by \cite{md-maenggi}, we adopt the meta distribution of the SINR as the performance metric, which is defined as the complementary cumulative distribution function (CCDF) of the conditional STP and answers the question ``What fraction of users can achieve the conditional STP value of at least $y$ (an arbitrary percentage value)?'' Recent works have focused on performance evaluation under the meta distribution framework in various wireless systems, including the heterogeneous networks \cite{md-hetnet}, D2D communications \cite{md-D2D}\cite{Deng-fine-grained}, coexisting sub-6GHz and mmWave networks \cite{md-hybrid}, coordinated multipoint transmission \cite{md-bs-cooperation}, non-orthogonal multi-access \cite{md-NOMA} and fractional power control \cite{md-power-control}.

In this paper, we develop a meta distribution-based framework for spatio-temporal analysis of the mmWave heterogeneous networks by utilizing stochastic geometry and queueing theory. This paper's contributions are summarized as follows.
\begin{enumerate}
\item The moments and the variance of the conditional STP are derived when the user is associated with a LOS/NLOS BS in each tier by taking the queue status into consideration. In addition, an equation is formulated to derive the analytical expressions of the meta distributions and a recursive approach is adopted to obtain a tight approximation of the meta distribution. The upper and lower bounds of the meta distribution are obtained in the second-degree dominant and favorable systems.
\item The numerical results show the effect of the key network parameters, i.e., blockage parameter, bias factor, number of antenna elements and density, on the network performance. The effect of the bernoulli rate on the active probability of the BSs under different blockage parameters is presented.
\end{enumerate}

\section{System Model}
We consider a downlink scenario in the mmWave heterogeneous networks consisting of $K$ tiers. Denote $\mathcal{K}\triangleq\{1,2,\cdots,K\}$. The BSs in each tier $k\in\mathcal{K}$ are assumed to be distributed as a homogeneous PPP $\Phi_k$ with density $\lambda_k$. The BSs in the $k$th tier transmit with power $P_k$. $\Phi_k,k\in\mathcal{K}$ are assumed to be independent and we denote by $\Phi$ the superposition of $\Phi_k,k\in\mathcal{K}$ as follows:
\begin{equation}
\Phi\triangleq\cup\Phi_k,k\in\mathcal{K}.
\end{equation}

We assume that the BSs in all tiers operate over the same mmWave frequency band and the bandwidth is $W$. Without loss of generality, according to Slivnyak's theorem \cite{slivnyak}, we study the performance of the typical user $u_0$ located at the origin.

The time is divided into discrete time slots with equal durations. For each user located within the Voronoi cell of its associated BS in $\Phi$, the arrival process of the packets is modeled as an independent Bernoulli process with parameter $\xi\in[0,1]$, which indicates that the packets for an arbitrary user arrive with probability $\xi$ in any time slot. Therefore, the queueing system for each user is considered as a Geo/G/1 queue. The sizes of all packets are assumed to be equal and one time slot is required to transmit a packet. In addition, an infinite-size buffer is allocated by each BS to store the incoming packets for each user located within its coverage. Note that multiple users can be served by a BS. In order to avoid contention between different users served by the same BS, we adopt the random scheduling scheme. Specifically, the BS randomly chooses a target user whose queue is non-empty in each time slot and transmits the 1st packet in the buffer to the corresponding user.

The wireless channel in the mmWave heterogeneous networks is characterized by the large-scale and small-scale fading. The BSs can be LOS or NLOS based on whether there is blockage intercepting the link between $u_0$ and the BS. The probability of a link between $u_0$ and a BS located at distance $x$ in the $k$th tier being LOS is determined by the LOS probability function $p_L(x)=e^{-\beta_kx}$, where $\beta_k$ is the blockage parameter for the $k$th tier and determined by the average size and density of the blockages. To distinguish the LOS/NLOS state of an arbitrary link, different path loss laws are employed for the LOS and NLOS links \cite{K-tier}:
\begin{equation}
L(x)=
\begin{cases}
\kappa_Lx^{\alpha_L}\ \ \ \text{with probability}\ p_L(x)\\
\kappa_Nx^{\alpha_N}\ \ \ \text{with probability}\ p_N(x),
\end{cases}
\end{equation}
where $\kappa_L$ and $\kappa_N$ are the intercepts for the LOS and NLOS link at 1 meter, while $\alpha_L$ and $\alpha_N$ the path loss exponents for the LOS and NLOS links.

For the small-scale fading, we denote by $h_{k,i}$ the small-scale fading term of link $i$ in the $k$th tier and assume Nakagami fading with the probability density function (PDF) $f(h)=2m_{\rho}^{m_{\rho}}h^{2m_{\rho}-1}e^{-m_{\rho}h^2}/\Gamma(m_{\rho}), \rho\in\{L,N\}$ for each link. Specifically, the Nakagami fadings with parameters $M_L$ and $M_N$ are applied to the LOS and NLOS links, respectively. Therefore, $|h_{k,i}|^2$ is a Gamma distributed random variable. For analytical tractability, we utilize a sector model to approximate the array antenna pattern as in \cite{Deng-fine-grained}. The beam direction of the interfering BSs is uniformly distributed over $[0,2\pi]$. The antenna array gain between $u_0$ and an interfering BS is given by
\begin{equation}
G=
\begin{cases}
G_{\max}, &\text{if}\quad |\varphi|\leq\frac{\delta}{2}\\
G_{\min}, &\text{otherwise},
\end{cases}
\end{equation}
where $G_{\max}$ denotes the main lobe gain, $G_{\min}$ the side lobe gain, $\varphi\in[-\pi,\pi)$ the angle of boresight direction and $\delta$ the half power beamwidth. We consider a $\sqrt{N}\times\sqrt{N}$ uniform planar array with $N$ elements, where $G_{\max}=N$, $G_{\min}=1/\sin^2\left(3\pi/2\sqrt{N}\right)$, and $\delta=\sqrt{3}/\sqrt{N}$ \cite{Deng-fine-grained}. A perfect alignment is assumed between $u_0$ and its serving BS and the maximum array gain $G_{\max}$ can be achieved for the link between $u_0$ and its serving BS.

Assume that $u_0$ is associated with the BS providing the strongest signal power among the BSs of all tiers. In addition, cell range expansion (CRE) may be adopted to offload some users from the Macro BSs (MBSs) to the small BSs (SBSs) in order to alleviate the burden of the MBSs and enhance the overall performance of the heterogeneous networks \cite{K-tier}. Therefore, $u_0$ should be associated with a BS based on the maximum biased received signal strength, which can be mathematically described
\begin{equation}
P_kB_kG_kL_k(x)^{-1}>P_jB_jG_jL_{j,\min}(x)^{-1},
\end{equation}
where $B_k$ denotes the bias factor of the $k$th tier, $L_{k,\min}(x)$ the minimum path loss between $u_0$ and the BSs in the $k$th tier.

A transmission is considered to be successful if the SINR at $u_0$ exceeds a predefined threshold $\theta$ and the feedback of the transmission, i.e., success or failure, can be detected by the BS immediately. If the transmission succeeds, the packet will be removed from the queue. Otherwise, the packet will be kept at the head of the corresponding queue and wait to be retransmitted. Therefore, whether a BS is transmitting or not depends on the queue status and the scheduling scheme at time slot $t$. Let $\zeta_{x,t,k}\in\{0,1\}$ be the indicator showing whether the BS at $x\in\Phi_k$ is transmitting a packet ($\zeta_{x,t,k}=1$) or not ($\zeta_{x,t,k}=0$) at time slot $t$. The SINR at $u_0$ is given by
\begin{equation}
\text{SINR}=\frac{P_kG_0|h_{k,0}|^2L_k(x)^{-1}}{\sigma^2+\sum_{j=1}^{K}\sum_{i\in\Phi_j\backslash B_{k,0}}\zeta_{x,t,j}P_jG_{j,i}|h_{j,i}|^2L_{j,i}^{-1}(x)},
\end{equation}
where $G_{j,i}$, $h_{j,i}$ and $L_{j,i}$ and $\sigma^2$ denote array gain, small-scale fading and the path loss for the interfering link, and the thermal noise power, respectively.

In order to shed light on the effect of the queue status on the network performance, we first assume that
\begin{equation}
\mathcal{P}\left(\zeta_{x,t,k}=1\right)=q_{a,k},
\end{equation}
and then obtain the analytical expression of the meta distribution by taking into account the temporal randomness of traffic.

Denoting the conditional STP
\begin{equation}
\mathcal{P}(\theta|\Phi)=\mathbbm{P}(\text{SINR}>\theta|\Phi),
\end{equation}
which is conditioned on the realization of $\Phi$, the SINR meta distribution is defined as the CCDF of $\mathbbm{P}(\theta|\Phi)$:
\begin{equation}
\bar{F}_{\mathcal{P}}(y)\triangleq\mathbbm{P}(\mathcal{P}(\theta|\Phi)>y),\ y\in[0,1].
\end{equation}

Due to the ergodicity of the point processes, the meta distribution can be regarded as the fraction of active links with their conditional STP greater than $y$.

\section{Auxiliary Results}
In this section, we first provide the characteristics of the path loss, i.e., the PDF and the complementary cumulative distribution function (CCDF) of the path loss, then provide the expression of the association probability. Without considering the LOS/NLOS state of the links, the CCDF and PDF of the path loss between $u_0$ and its serving BS are given in the following lemma.

\begin{Lemma}
The CCDF of the path loss between $u_0$ and its serving LOS/NLOS BS in the $k$th tier is given by
\begin{equation}\label{path-loss-CCDF}
\begin{split}
\bar{F}_{L_{k,\rho}}(x)=\exp(-\Lambda_{k,\rho}([0,x))),k\in\mathcal{K}, \rho\in\{L,N\},
\end{split}
\end{equation}
where
\begin{equation}\label{LOS-intensity}
\begin{split}
&\Lambda_{k,L}([0,x))\\
&=2\pi\lambda_k\beta_k^{-2}\left(1-e^{-\beta_k(x/\kappa_L)^{1/\alpha_L}}
\left(1+\beta_k(x/\kappa_L)^{1/\alpha_L}\right)\right),
\end{split}
\end{equation}

\begin{equation}\label{NLOS-intensity}
\begin{split}
&\Lambda_{k,N}([0,x))=\pi\lambda_k(x/\kappa_N)^{2/\alpha_N}\\
&-2\pi\lambda_k\beta_k^{-2}\left(1-e^{-\beta_k(x/\kappa_N)^{1/\alpha_N}}
\left(1+\beta_k(x/\kappa_N)^{1/\alpha_N}\right)\right).
\end{split}
\end{equation}
\end{Lemma}

\emph{Proof:} Please refer to \cite{journal-version}.

Next, we obtain the expression of the association probability in the following lemma.
\begin{Lemma}\label{lemma-association}
The probability that $u_0$ is associated with the BS in the $k$th tier is given by
\begin{equation}
A_{k}=\int_{0}^{\infty}\Lambda_{k}'([0,l_k))e^{\sum_{j=1}^{K}\Lambda_{j}\left([0,\frac{P_jB_JG_j}{P_kB_kG_k}l_k)\right)}\text{d}l_k,
\end{equation}
where $\Lambda_{k}'([0,l_k))$ is the derivative of $\Lambda_{k}([0,l_k))$.
\end{Lemma}

\emph{Proof:} Please refer to \cite{journal-version}.

From Lemma \ref{lemma-association}, we observe that the association probability is dependent on three sets of network parameters, i.e., the deployment parameters, the antenna array parameters and the mmWave environment parameters. The deployment parameters include the transmit power $P$ and the BS density $\lambda_k$. The antenna array parameters consist of the main lobe gain $G_{\max}$, the side lobe gain $G_{\min}$ and the half power beamwidth $\delta$. The mmWave environment parameters consist of the blockage parameters $\beta_k$ and the path loss exponent $\alpha_{\rho},\rho\in\{L,N\}$.

\section{Analysis of Meta Distribution}
In this section, we first provide the analytical expression of the moments of the conditional STP. Then the expression of the meta distribution of the SINR distribution is provided. In addition, the approximation of the moments of the conditional STP is derived.

\begin{Theorem}\label{theorem-moment}
Given that $u_0$ is associated with a LOS/NLOS BS in the $k$th tier, the $b$-th moment of the conditional STP is given by
\begin{equation}\label{equation-moment}
\begin{split}
&M_{b,k,\rho}=\frac{1}{A_{k,\rho}}\int_{0}^{\infty}f_{L_{k,\rho}}(l_{k,\rho})\left(\sum_{\tau_1=0}^{\infty}\sum_{\tau_2=0}^{m_{\rho}\tau_1}\binom{b}{\tau_1}\binom{m_{\rho}\tau_1}{\tau_2}
\right.\\
&\left.(-1)^{\tau_1+\tau_2}e^{-s\sigma^2\zeta_{\rho}\tau_2}\prod_{j=1}^{K}\mathcal{L}_{I_{j,L}}(s\zeta_{\rho}\tau_2)\mathcal{L}_{I_{j,N}}(s\zeta_{\rho}\tau_2)\right)\text{d}l_{k,\rho},
\end{split}
\end{equation}
where $\zeta_{\rho}=(m_{\rho}!)^{-1/m_{\rho}}$, $\mathcal{L}_{I_{j,\upsilon}}(s\zeta_{\rho}\tau_2),\upsilon\in\{L,N\}$ is given by
\begin{equation}\label{Laplace-transform}
\begin{split}
&\mathcal{L}_{I_{j,\upsilon}}(s\zeta_{\rho}\tau_2)=\exp\left(\int_{\frac{P_jB_j}{P_kB_k}l_{k,\rho}}^{\infty}\right.\\
&\left.\left(1-\left(1+\frac{\theta L_{k,\rho}m_{\rho}P_j\zeta_{\rho}\tau_2}{xP_kG_0M_{\upsilon}}\right)^{-M_{\upsilon}}\right)q_{a,k}\Lambda_{j,\upsilon}(\text{d}x)\right).
\end{split}
\end{equation}

\end{Theorem}
\emph{Proof:} See Appendix A.

By utilizing the law of total probability, the $b$-th moment of the conditional STP for the $K$-tier networks is given by
\begin{equation}
M_{b}=\sum_{k=1}^{K}\sum_{\rho\in\{L,N\}}A_{k,\rho}M_{b,k,\rho}.
\end{equation}

By applying the Gil-Pelaez theorem \cite{Gil-Pelaez}, the meta distribution of the SINR is given by
\begin{equation}\label{md}
\bar{F}_{\mathcal{P}}(y)=\frac{1}{2}+\frac{1}{\pi}\int_{0}^{\infty}\frac{\mathcal{J}\left(e^{-jt\log y}M_{jt}\right)}{t}\text{d}t,
\end{equation}
where $\mathcal{J}(z)$ is the imaginary part of $z$. Since the numerical evaluation of (\ref{md}) is cumbersome and it is difficult to obtain further insight, we utilize a beta distribution to approximate the meta distribution by matching the first and second moments, which can be easily obtained from the result in (\ref{equation-moment}).

By matching the variance and mean of the beta distribution, i.e., $M_{2}-M_{1}^2$ and $M_{1}$, the approximated meta distribution of the SINR can be given by
\begin{equation}\label{md-approximation}
\bar{F}_{\mathcal{P}_{k,\rho}}\approx 1-I_y\left(\frac{M_{1}\beta}{1-M_{1}},\beta\right),\ y\in[0,1],
\end{equation}
where
\begin{equation}
\beta=\frac{(M_{1}-M_{2})(1-M_{1})}{M_{2}-M_{1}^2},
\end{equation}
and $I_y(a,b)$ is the regularized incomplete beta function
\begin{equation}
I_y(a,b)\triangleq\frac{\int_{0}^{y}t^{a-1}(1-t)^{b-1}\text{d}t}{\text{B}(a,b)}.
\end{equation}

Note that the randomness of the networks is from two components, i.e., the random locations of the BSs and the corresponding active states. In practice, $q_{a,k}$ is closely related to the queue status, which is jointly affected by the conditional STP for the $k$th tier and the arrival rate. We denote by $\nu_k$ the number of users served by each BS in the $k$th tier. Given that $u_0$ is associated with the BS in the $k$th tier, the probability that the packet is successfully transmitted from the BS in the $k$th tier to $u_0$ in a time slot equals the product of the conditional STP for the $k$th tier and the probability that $u_0$ is scheduled, i.e., $\mathcal{P}_k(\theta|\Phi)/\nu_k$, which is also named the service rate. Therefore, the active probability of the BS is
\begin{equation}\label{active-probability}
q_{a,k}=
\begin{cases}
1, &\text{if}\ \mathcal{P}_k(\theta|\Phi)\leq\nu_k\xi\\
\frac{\nu_k\xi}{\mathcal{P}_k(\theta|\Phi)}, &\text{if}\ \mathcal{P}_k(\theta|\Phi)>\nu_k\xi
\end{cases}
\end{equation}
\begin{Remark}
Recall that the queueing process for the typical user is considered as a Geo/G/1 process. According to whether the conditional STP of the corresponding link is larger than the arrival rate, we can divide the BS into two sets, i.e., the first set with the conditional STP larger than the arrival rate and the second set with the conditional STP lower than the arrival rate. By utilizing the meta distribution of the SINR, the fractions of two sets can be characterized, i.e., the first set accounts for $\bar{F}_{\mathcal{P}_k}(\xi)$ and the second set accounts for $F_{\mathcal{P}_k}(\nu\xi)$ of all BSs with users located in their coverage. For the queues with the service rate larger than the arrival rate, the non-empty probability equals the utilization of the queue $\xi/\mathcal{P}_k(\theta|\Phi)$. Therefore, the active probability for the BSs in the first set of the $k$th tier is $\xi/\mathcal{P}_k(\theta|\Phi)$. For the queues with the service rate lower than the arrival rate, the non-empty probability is 1 since the corresponding queues are backlogged and will never be empty. In other words, the BSs in the second set will always be active.
\end{Remark}

Next, we derive the mean active probability of each tier in the mmWave heterogeneous networks. In the mmWave heterogeneous networks, some BSs may have no users to serve and the active probability of the corresponding link is zero. According to \cite{PMF}, the distribution of the number of users within the coverage of each BS in the $k$th tier is provided in the following lemma.
\begin{Lemma}
The probability mass function (PMF) of the number of the users served by the BS in the $k$th tier is
\begin{align}\label{PMF-equation}
&U_{k}(\nu)\\
&\triangleq\frac{1}{\Gamma(\nu)}\left(\frac{\lambda_u A_k x}{3.5\lambda_k}\right)^{\nu-1}\frac{\Gamma(3.5+\nu)}{\Gamma(4.5)}
\left(1+\frac{\lambda_u A_k x}{3.5\lambda_k}\right)^{-3.5-\nu}.
\end{align}
\end{Lemma}

Due to the ergodicity of the PPP, since the probability of a user being associated with the BS in the $k$th tier is $\mathcal{A}_k$, the average fraction of the users served by the BSs in the $k$th tier is also $\mathcal{A}_k$. Therefore, the density of users being associated with the BSs in the $k$th tier is $\lambda_u\mathcal{A}_k$. By replacing $\lambda_u$ with $\lambda_u\mathcal{A}_k$, we can obtain the result in (\ref{PMF-equation}).

The mean active probability for the BSs can be derived as
\begin{equation}\label{mean-active-probability}
\mathbbm{E}[q_{a,k}]=\sum_{\nu=1}^{\infty}\frac{U_{k}(\nu)}{1-U_{k}(0)}
\left(1-\bar{F}_{\mathcal{P}_k}(\nu\xi)-\int_{\nu\xi}^{1}\frac{\nu\xi}{s}\bar{F}_{\mathcal{P}_k}(\nu\xi)\right).
\end{equation}

Based on the above analysis, to obtain the meta distribution, we have the following lemma.
\begin{Lemma}
The meta distribution of the SINR for the $k$th tier of the mmWave heterogeneous networks is
\begin{equation}\label{md-fixed-point}
\begin{split}
&\bar{F}_{\mathcal{P}_k}=\frac{1}{2}+\frac{1}{\pi}\int_{0}^{\infty}\frac{1}{t}\mathcal{J}\left(e^{-jt\log x}
\sum_{k=1}^{K}\sum_{\rho\in\{L,N\}}\int_{0}^{\infty}f_{L_{k,\rho}}(l_{k,\rho})\right.\\
&\left(\sum_{\tau_1=0}^{\infty}\sum_{\tau_2=0}^{m_{\rho}\tau_1}\frac{\Gamma(jt+1)}{\Gamma(\tau_1+1)\Gamma(jt-\tau_1+1)}\binom{m_{\rho}\tau_1}{\tau_2}(-1)^{\tau_1+\tau_2}
\right.\\
&\left.\left.\left.e^{-s\sigma^2\zeta_{\rho}\tau_2}\prod_{j=1}^{K}\mathcal{L}_{I_{j,L}}(s\zeta_{\rho}\tau_2)\mathcal{L}_{I_{j,N}}(s\zeta_{\rho}\tau_2)\right)\text{d}l_{k,\rho}
\right.\right)\text{d}t
\end{split}
\end{equation}
where $\mathcal{L}_{I_{j,\upsilon}}(s\zeta_{\rho}\tau_2),\upsilon\in\{L,N\}$ is given by
\begin{equation}\label{Laplace-transform-lemma}
\begin{split}
&\mathcal{L}_{I_{j,\upsilon}}(s\zeta_{\rho}\tau_2)=\exp\left(\sum_{\nu=1}^{\infty}\frac{U_{k}(\nu)}{1-U_{k}(0)}\int_{\frac{P_jB_j}{P_kB_k}l_{k,\rho}}^{\infty}\left(1-\left(1+\right.\right.\right.\\
&\left.\left.\left.\frac{\theta L_{k,\rho}m_{\rho}P_j\zeta_{\rho}\tau_2}{xP_kG_0M_{\upsilon}}\right)^{-M_{\upsilon}}\right)\left(1-\bar{F}_{\mathcal{P}_k}(\nu\xi)-\int_{\nu\xi}^{1}\frac{\nu\xi}{s}\bar{F}_{\mathcal{P}_k}(\nu\xi)\right)\right.\\
&\left.\Lambda_{j,\upsilon}(\text{d}x)\right).
\end{split}
\end{equation}
\end{Lemma}

The solution of the equation (\ref{md-fixed-point}) is the SINR meta distribution of the mmWave heterogeneous networks. A recursive approach is proposed to obtain a tight approximation of the meta distribution. Since we temporal correlation of the transmission success events among different time slots can be reduced significantly with random scheduling, we assume that the transmission success events among different time slots are independent. With the recursive approach, the meta distribution is
\begin{equation}
\bar{F}_{\mathcal{P}_k}(x)=\lim_{t\rightarrow\infty}\bar{F}_{\mathcal{P}_{k,t}}(x),k\in\mathcal{K}.
\end{equation}

In the recursive process, the active probability for each tier is obtained iteratively until a stationary solution is reached. When the active probability of the BSs in all tiers other than $k$ is fixed, we use the meta distribution of the SINR for the $k$th tier at time slot $t-1$ to characterize the active probability of the BSs in the $k$th tier at time slot $t$. Since all the queues are empty at time slot 0, the probability of the corresponding queues being non-empty is equal to the arrival rate $\nu\xi$. Therefore, the initial value of the active probability is $\nu\xi$. A tight approximation can be obtained when $t\rightarrow\infty$.

\section{Numerical Results}
In this section, we consider a two-tier mmWave heterogeneous network, where the SBS (Tier $k=2$) is overlaid with the MBSs (Tier $k=1$). We first present the comparison between the impact of the SINR threshold on the STP with/without queueing, and then present the relationship between the SBS active probability and the Bernoulli parameter. Unless otherwise stated, the parameters are set as listed in the following table.

\newcommand{\tabincell}[2]{\begin{tabular}{@{}#1@{}}#2\end{tabular}}  
\begin{table}[htbp]
\centering
\caption{\label{tab:test}System Parameters}
\begin{tabular}{|c|c|c|}
\hline
$\textbf{Parameters}$&$\textbf{Values}$\\
\hline
Bernoulli parameter & \tabincell{c}{$\xi=0.3$} \\
\hline
Transmit power & \tabincell{c}{$P_1=43$dBm, $P_2=23$dBm} \\
\hline
Bias factor & $B_1=1$, $B_2=1$\\
\hline
Path loss exponent & $\alpha_{L}=2$, $\alpha_{N}=4$ \\
\hline
Density & \tabincell{c}{$\lambda_1=5/(500^2\pi)$, $\lambda_2=10/(500^2\pi)$}\\
\hline
Blockage parameter & \tabincell{c}{$\beta_1=0.006$, $\beta_2=0.024$}\\
\hline
Nakagami parameter& $ m_{L}=3$, $m_{N}=2$\\
\hline
Bandwidth & \tabincell{c}{$W=1$G}\\
\hline
Carrier frequency & 28GHz\\
\hline
Path loss intercept & $\kappa_{L}=\kappa_{N}=(F_c/4\pi)^2$\\
\hline
\end{tabular}
\end{table}

Fig. 1 plots the STP as the function of the SINR threshold $\theta$. The simulation results match the theoretical analysis well. The performance fluctuation can be reflected by the variance of the conditional STP. A large variance corresponds to a large performance fluctuation and vice versa \cite{deng-SINR}. From Fig. 1, it can be observed that the STP decreases with the SINR. In addition, the STP for the networks with queueing is larger than that for the network with a full buffer assumption.
\begin{figure}
  \centering
  \includegraphics[width=3.5in]{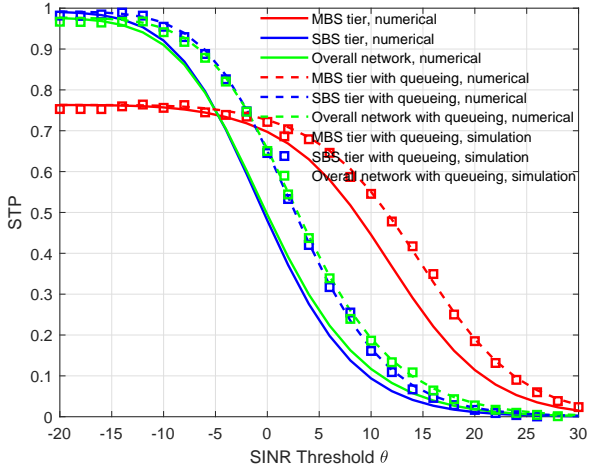}
  \caption{The impact of SINR threshold $\theta$ on the per-tier STP.}\label{sinr-threshold}
\end{figure}

Fig. 2 illustrates the impact of the arrival rate on the BS active probability. It is observed that the BS active probability increases monotonically when increasing the arrival rate. In addition, the BS active probability increases when the blockage parameter becomes larger. This phenomenon indicates that the BSs in the high blockage parameter regime are more likely to be affected by the traffic. This is because the STP decreases when the blockage parameter increases and the packets cannot be served in a timely fashion. As such, the queues are more likely to be backlogged and thus the active probability becomes larger.
\begin{figure}
  \centering
  \includegraphics[width=3.5in]{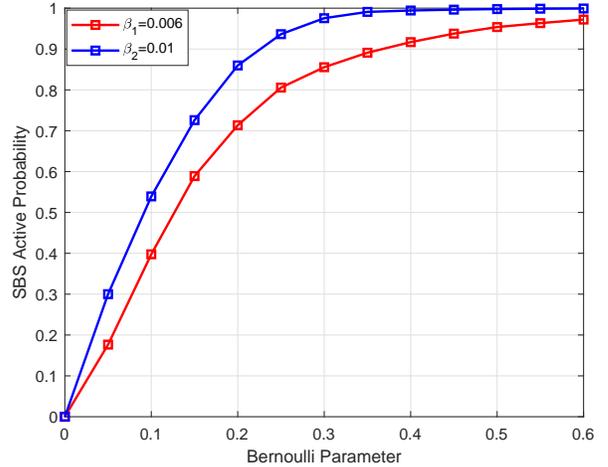}
  \caption{SBS Active probability as the function of Bernoulli parameter under different blockage parameters.}\label{active-arrival}
\end{figure}

\section{Conclusion}
We have focused on the meta distribution of the SINR in the mmWave heterogeneous networks. By utilizing stochastic geometry and queueing theory, an equation has been formulated to derive the meta distribution and the meta distribution can obtained in a recursive manner. Two tight bounds of the meta distribution, i.e., the second-degree dominant and favorable system, have been provided. In addition, the key performance metrics, i.e., the STP, the variance of the conditional STP and the mean delay, have been obtained. Finally, the numerical results have revealed the impact of the key network parameters, such as the blockage parameter and the SINR threshold, on the network performance.
\section{Appendix}
\subsection{Proof of Theorem \ref{theorem-moment}}
Given that $u_0$ is associated with a LOS/NLOS BS in the $k$th tier, the $b$-th moment of the conditional STP can be expressed as
\begin{equation}
\begin{split}
&M_{b,k,\rho}=\mathbbm{P}\left(\frac{P_kG_0h_{k,0}L_k^{-1}(x)}{\sigma^2+I}>\theta\right)^b\\
&\overset{(a)}{=}\mathbbm{E}_{I,s}\left[\left(1-\frac{\gamma(m_{\rho},s(\sigma^2+I))}{\Gamma(m_{\rho})}\right)^b\right]\\
&\overset{(b)}{\approx}\mathbbm{E}_{I,s}\left[\left(1-\left(1-e^{-s(\sigma^2+I)\zeta_{\rho}}\right)^{m_{\rho}}\right)^b\right]\\
&\overset{(c)}{=}\mathbbm{E}_{I,s}\left[\sum_{\tau_1=0}^{\infty}\binom{b}{\tau_1}\left(-\left(1-e^{-s(\sigma^2+I)\zeta_{\rho}}\right)^{m_{\rho}}\right)^{\tau_1}\right]\\
&\overset{(d)}{=}\mathbbm{E}_{I,s}\left[\sum_{\tau_1=0}^{\infty}\sum_{\tau_2=0}^{m_{\rho}\tau_1}\binom{b}{\tau_1}\binom{m_{\rho}\tau_1}{\tau_2}
(-1)^{\tau_1+\tau_2}\exp(-s\sigma^2\zeta_{\rho}\tau_2)\right.\\
&\left.\prod_{j=1}^{K}\mathcal{L}_{I_{j,L}}(s\zeta_{\rho}\tau_2)\mathcal{L}_{I_{j,N}}(s\zeta_{\rho}\tau_2)\right],
\end{split}
\end{equation}
where (a) follows from $\Gamma(s)=\gamma(s,x)+\Gamma(s,x)$, (b) is from that the CDF of a Gamma random variable can be tightly upper bounded by $\gamma\left(m_{\rho},s(I+\sigma^2)\right)/\Gamma(m_{\rho})<[1-e^{-s(I+\sigma^2)\zeta_{\rho}}]$, (c) and (d) follows from the binomial expansion theorem, the definition of the Laplace transform of the interference plus noise and $\mathcal{L}(s)=\exp(-s\sigma^2)\prod_{j=1}^{K}\mathcal{L}_{I_{j,L}}(s)\mathcal{L}_{I_{j,N}}(s)$. Then the proof of Theorem \ref{theorem-moment} can be completed by averaging over $L_{k,\rho}$.

\end{document}